\documentclass{article}

\usepackage{arxiv}

\usepackage[utf8]{inputenc}
\usepackage[T1]{fontenc}

\usepackage{float}

\usepackage{tabularx}
\usepackage{tikz}
\usetikzlibrary{positioning}
\usepackage{booktabs}

\usepackage{xurl}
\usepackage[hidelinks]{hyperref}
\usepackage{graphicx}

\usepackage[numbers]{natbib}
\title{From Transactions to Records: Reconceptualizing Blockchain Systems through a Lifecycle Lens}

\author{
Tom Barbereau\\
Netherlands Institute for Applied Scientific Research (TNO)\\
The Hague, The Netherlands\\
University of Amsterdam\\
Amsterdam, The Netherlands\\
\texttt{tom.barbereau@tno.nl}
\And
Ruggero Montalto\\
Netherlands Institute for Applied Scientific Research (TNO)\\
The Hague, The Netherlands
\And
Christian Beyer\\
University of Applied Sciences for Police and Public Administration\\
North Rhine-Westphalia, Germany
}

\begin{document}

\maketitle

\begin{abstract}
Current blockchain research and analytics tend to prioritize observable on-chain transactions, obscuring the processes through which cryptocurrencies are created, publicised, retained, and disposed of.
In response, this paper considers distributed ledger technologies from records management principles in ISO 15489-1:2016. Setting off by specifying the parallels --- that is transactions as ``records'', crypto-asset units as ``information assets'', and blockchains as ``aggregations'' --- we introduce a seven-stage lifecycle for blockchain data.
We apply the framework to Bitcoin, a fungible token, and a non-fungible token. On this basis, we argue that blockchain systems are not merely transactional infrastructures but record management systems with distinctive characteristics.
We discuss how the on-chain/off-chain boundary and privacy-enhancing technologies can complicate lifecycle visibility, with particular relevance for crypto-crime research and investigation. As a meta-level framework, the lifecycle perspective enables positioning existing research, decomposing legal, regulatory, technological, and operational challenges by stage, and informing lifecycle-aware approaches to blockchain governance, analytics, and regulation.
\end{abstract}


\section{Introduction}

Existing research on blockchain -- particularly that which is concerned with targeting illicit or suspicious activity -- largely focuses on observable on-chain transactions. These works address only the visible tip of the iceberg, limited to what is colloquially referred to as `following the money' \citep{Dearden2023Mar}. Though improved methods of tracing have reduced crime in the space by giving evidential value to law enforcement agencies and prosecutors \citep{Frowis2020Jun}, this transaction- and usage-centric focus obscures the underlying informational processes through which data in blockchain systems is created, publicised, retained, and possibly disposed of. This is the hidden side of the iceberg. As a result, current approaches capture illicit behaviour without adequately accounting for the broader structures that enable, constrain, or conceal such activity \citep{Wright2024Feb}. Importantly, these deeper layers are neither unknown nor inaccessible: established frameworks for document and records lifecycle management already offer systematic ways that point at how information assets evolve over time.

The lifecycle of documents and records has remained conceptually stable across technological contexts, from physical records to digital information systems. ISO 15489-1:2016 (Information and Documentation – Records Management) \citep{iso15489} provides clear definitions of records, information assets, and authoritative sources. Yet, despite crypto-assets being fundamentally data-based, the perspective of records lifecycle management has not been applied here. 
Thus far, efforts have largely been on tracing usage through heuristic and statistical methods \citep{Agarwal2024Mar,DUDANI2023301576}, identifying possible system vulnerabilities \citep{chan2020,Zhang2019Jul}, providing taxonomies and classifications of certain tools or characteristics \citep{barbereau2023beyond,Hartwich2024Feb}, as well as addressing externalities in the form of judicial questions \citep{DeFilippi2022Sep,pelker2021using} or the effects of regulations \citep{Bodo2025regulation,Feinstein2021May}. 
Each of these largely ignores the reality that crypto-assets are mere ``information assets'' that are managed by a ``record'' of transactions that are ``aggregated'' in a blockchain. 

In response to this shortcoming, setting off from ISO 15489-1:2016, this paper puts forward a staged lifecycle for blockchain data. Section \ref{sec:background} provides background on records lifecycle management, formalising the lifecycle in terms of seven stages. Section \ref{sec:method} presents the research design, which consists of three phases. Section \ref{sec:analysis} presents the analysis corresponding to the three research phases. In the specification phase, we formalise blockchain elements through ISO 15489-1:2016. Second, in the conceptualisation phase, we take the seven-stage lifecycle and formalise it for blockchain data. This represents our core theoretical contribution. Third, in the application phase based on a multiple case study \citep{yin2009case}, we apply this framework to UTXO-based systems (Bitcoin) and EVM-based systems, including both fungible (Tether) and non-fungible tokens. Section \ref{sec:discussion} presents implications for theory and practice. Notably, the proposed framework challenges dominant blockchain scholarship by rejecting assumptions of linear, stable stages, instead emphasizing entanglement, feedback loops, and shifting meanings. By foregrounding longitudinal trajectories and end-of-life considerations (i.e. forms of disposition), it provides a unifying structure that reorients both theory and practice toward responsibility, temporality, and stage-specific intervention rather than static or use-focused analyses. 

%The paper is grounded in systems-theoretical and cybernetic perspectives on self-organization \citep{vonFoerster1960}, with particular emphasis on the role of information and data flows in the emergence and maintenance of system order. It is rooted firmly in the literature on document lifecycle management \citep{Kwon1987Apr,nolan1974managing}. Complementary to that, it draws on and synthesizes insights from computer science research, information systems scholarship, information law, and literature on blockchain governance more broadly. Methodologically, the paper considers three cases as source of inductive evidence \citep{yin2009case}: Bitcoin (representative of the UTXO-managed lifecycle), Tether and a (fictionalised) non-fungible token (representative of an account-managed lifecycle). 
%By nature, this commentary is thereby conceptual-empirical \citep{mora2008conceptual,Jaakkola2020conceptual}. 

\section{Background}
%\subsection{Documents, data, and lifecycles}
\label{sec:background}

Though records lifecycle management is hardly a science -- a statement corroborated by the objective scarcity of scientific literature around the topic \citep{Rahul2020Jan} -- its perspectives are widely applied and valued in practice. That practical value is the motivation for our work. 
Most prominent is the international standard ISO 15489-1 \citep{iso15489}. 
Section 5.1 of the standard defines records as ``both evidence of business activity and information assets'', stating that ``any set of information, regardless of its structure or form, can be managed as a record''. Section 5.2.1 stipulates that ``records document individual events or transactions, or may form aggregations that have been designed to document work processes, activities or functions'', and that ``records, regardless of form or structure, should possess the characteristics of authenticity, reliability, integrity and usability [...] to be considered authoritative evidence of business events or transactions and to fully meet the requirements of the business''. 
Authenticity requires proof that a record is what it claims to be, was created or sent by the stated agent, and was created or sent at the stated time. Reliability requires that a record's contents fully and accurately represent the transactions, activities, or facts to which they attest, and that the record can be depended upon for subsequent use. Integrity means that a record is complete and unaltered. Usability means that a record can be located, retrieved, presented, and interpreted within a reasonable time-frame. 
Each of these four characteristics is discussed in greater detail in the literature \citep[refer to][]{Kastenhofer2015Sep}.

To ensure these characteristics of authoritative information are valid throughout all stages of a document's lifecycle, records management systems comprise (business) rules, processes, policies, and procedures \citep{Kastenhofer2015Sep}. These systems often enforce many controls, of varying nature (technical, legal, managerial, or administrative) to ensure authenticity (i.e. record creators must be authorized and identified), reliability (i.e. records should be created at the time of the event, or soon after, by individuals with direct knowledge of the facts or by systems routinely used to conduct the transaction), integrity (i.e. records must be protected against unauthorized alteration), and usability (i.e. records must be connected to the business processes or transactions that produced them, and linkages between related records must be maintained).

\begin{table*}
\caption{Overview of records management lifecycle stages}
\label{tab:data-lifecycle}
\centering
\small
\begin{tabularx}{\textwidth}{X X X X X}
\toprule
US Geological Survey \citep{Faundeen2014} & \citet{harvard2025} & \citet{McElhone2022} & \citet{Rahul2020Jan} & \citet{wisconsin2022} \\
\midrule
Plan      & Plan \& design           &            &            & Plan    \\
Acquire   & Collect \& create        & Create     & Create     & Create  \\
Process   &                          & Store      & Store      &         \\
Analyse   & Analyse \& collaborate   & Use        & Use        & Use     \\
Preserve  & Evaluate \& archive      & Archive    & Archive    &         \\
          & Share \& disseminate     & Share      & Share      & Share   \\
Publish/share & Publish \& reuse     &            &            &         \\
          &                          & Destroy    & Destroy    & Destroy \\
\bottomrule
\end{tabularx}
\end{table*}

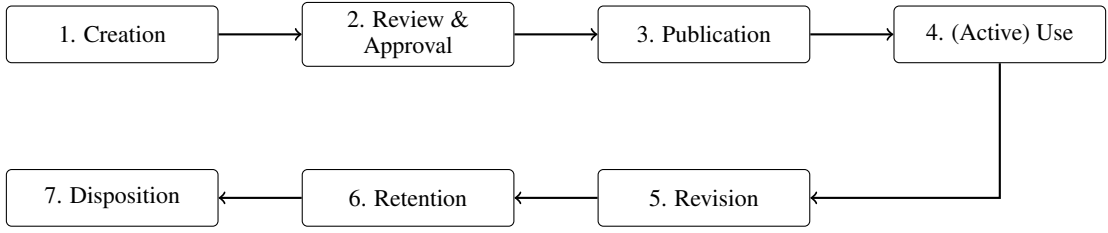
\begin{figure*}[t]
\centering
\begin{tikzpicture}[
    node distance=1.4cm and 1.1cm,
    every node/.style={
        rectangle,
        rounded corners=2pt,
        draw,
        align=center,
        font=\small,
        minimum width=2.8cm,
        minimum height=0.75cm,
        inner sep=3pt
    },
    arrow/.style={->, thick}
]

\node (creation) {1. Creation};
\node (review) [right=of creation] {2. Review \&\\ Approval};
\node (publication) [right=of review] {3. Publication};
\node (use) [right=of publication] {4. (Active) Use};

\node (revision) [below=of publication] {5. Revision};
\node (retention) [left=of revision] {6. Retention};
\node (disposition) [left=of retention] {7. Disposition};

\draw[arrow] (creation) -- (review);
\draw[arrow] (review) -- (publication);
\draw[arrow] (publication) -- (use);
\draw[arrow] (use.south) |- (revision.east);
\draw[arrow] (revision) -- (retention);
\draw[arrow] (retention) -- (disposition);

\end{tikzpicture}
\caption{Seven-stage document lifecycle}
\label{fig:sevenstages}
\end{figure*}

There are numerous records management systems that formalise stages of a document's lifecycle, each encompassing several steps (Table \ref{tab:data-lifecycle}). Against this backdrop, and considering the scope of our argument, we postulate a seven-stage lifecycle (Figure \ref{fig:sevenstages}):

\begin{enumerate}
    \item \textbf{Creation}: We produce accurate, well-structured documents (with proper metadata) for future retrieval and management.
    \item \textbf{Review \& Approval}: We ensure document accuracy, legal validity, compliance, and alignment with (organizational) standards before publication.
    \item \textbf{Publication}: We make approved documents available to authorized users through appropriate channels.
    \item \textbf{(Active) Use}: We enable efficient access and use of documents.
    \item \textbf{Revision}: We update documents to reflect current information, while maintaining version history.
    \item \textbf{Retention}: We maintain documents for required retention periods according to legal, regulatory, and business requirements.
    \item \textbf{Disposition}: We properly retire documents that have reached end of retention period through archival or secure destruction.
\end{enumerate}

As an example familiar to most, let us apply the postulated lifecycle and its seven stages to a national passport:

\begin{enumerate}
    \item \textbf{Creation}: A citizen applies for a passport. The government agency captures biometric data, photographs, and personal information, creating a structured digital record with metadata (applicant ID, date, issuing office) to be printed into the physical passport booklet.
    \item \textbf{Review \& Approval}: Immigration officers verify the applicant's identity documents, check for criminal records, confirm citizenship status, and ensure all information meets legal standards and international passport regulations (ICAO standards \citep{Juneja2025Jun}) before approving issuance.
    \item \textbf{Publication}: The approved passport is printed with security features, laminated, and officially issued to the citizen. It is registered in the national passport database and made 'active' for international travel use.
    \item \textbf{Active Use}: The citizen uses the passport for travel. Border control officers scan it at airports, adding entry/exit stamps. The passport number links to visa applications, and customs systems access the document's validity status in real time.
    \item \textbf{Revision}: The passport holder changes their name due to marriage. They apply for a passport update, and a new version is issued with the updated name whilst maintaining a record linking it to the previous passport number in the system.
    \item \textbf{Retention}: After the ten-year validity period expires, the physical passport is returned to the holder (marked as expired), but the government retains digital records of the passport's issuance and travel history for the legally required period (often 15–25 years) for security and immigration purposes.
    \item \textbf{Disposition}: After the retention period ends, digital records are either archived in a permanent historical database (for statistical purposes, with personal identifiers removed) or securely deleted according to data protection laws, whilst expired physical passports are destroyed by the holder or collected and shredded by authorities.
\end{enumerate}

\section{Method}
\label{sec:method}
We followed a qualitative, theory-driven research method. The analysis combines an application of an established theoretical framework with empirical illustration through case analysis \citep{yin2009case}. Empirical material consists of publicly available technical documentation, protocol specifications, regulatory texts, and secondary sources, which were analysed to examine how crypto-asset lifecycles unfold across different blockchain types. This approach enables analytic rather than statistical generalization, grounding theoretical insights in empirically observable characteristics and nuances of cases \citep{Benbasat1987Sep,Darke1998Oct}.
Our analysis proceeds in three phases: 

\paragraph{1. Specification.} First, blockchain elements are analytically re-specified through a records management lens (as specified in ISO 15489-1:2016 \citep{iso15489}): transactions are treated as ``records'', crypto-assets as ``information assets'', and blockchains as ``aggregations'' that collectively maintain system state. This re-specification establishes conceptual equivalence between established information management constructs and blockchain artefacts. If blockchain data are documents, then the document lifecycle management perspective is applicable. 

\paragraph{2. Conceptualisation.} Second, the seven stages derived from records management (refer to Section \ref{sec:background}) are systematically mapped onto blockchain data. This mapping is used to conceptually examine how different technical architectures and governance arrangements shape lifecycle progression. Importantly, the analysis does not presume linear or clean transitions between stages; instead, it allows for overlap, recursion, and persistence, reflecting the operational characteristics of real-world systems and the properties of blockchain data.

\paragraph{3. Application.} The conceptually derived seven-stage lifecycle for blockchain data is applied empirically to three cases, which serve as analytically contrasting instantiations of different blockchain types \citep{yin2009case}. The two predominant types are Unspent Transaction Output (UTXO) and account-based blockchains \citep{Liu2022Nov,Akcora2022Jan}. Bitcoin is examined as a representative case of a UTXO-managed blockchain,\footnote{In a UTXO-based ledger, one's balance is not a single stored number but the sum of discrete unspent coin amounts, each representing an UTXO. When one sends funds, existing UTXOs are consumed as inputs and new ones are produced in their place. The network prevents double-spending by verifying that inputs exist and have not been spent before. A wallet's total balance is therefore a software abstraction, the sum of all UTXOs assigned to one's addresses.} in which system state is maintained through discrete, spendable outputs. Tether and a fictitious non-fungible token are analysed as representative of account-managed blockchains, where balances or ownership states are updated through account-based mechanisms. Together, they provide sufficient diversity \citep{yin2009case} in their representation of the problem space \citep{Benbasat1987Sep,Darke1998Oct}. In sum, through this comparative application, the cases provide inductive grounding for the lifecycle framework and illustrate how different architectural and governance choices give rise to distinct dynamics.

To operationalize this analysis, we integrate complementary insights from multiple bodies of literature \citep{VomBrocke2015}. Computer science research informs the treatment of data structures, protocol-level state management, and transaction processing; information systems scholarship contributes perspectives on digital artifacts as socio-technical constructs embedded in institutional settings; and, information law provides grounding for questions of control, responsibility, and retention of data, while the blockchain governance literature informs the analysis of rule-setting, enforcement, and authority across lifecycle stages. 
\section{Analysis and Findings}
\label{sec:analysis}

\subsection{Specification: Crypto-assets as data in a lifecycle}

%This section treats cryptocurrency systems as data systems: transactions as records, cryptocurrency units as state, and blockchains as replicated, append-only ledgers. We map ISO 15489-1:2016 \citep{iso15489} terms onto these primitives.
%, and then test where the lifecycle analogy holds given decentralised governance, append-only updates, and on-chain persistence (Section~\ref{subsec:different}).

Prior to conducting a formal analysis of crypto-assets as records management system, there is a need for specification. ISO 15489 delineates managed data into \emph{records} (evidence of an event), \emph{information assets} (data that carries value), and \emph{aggregations} (structured collections of records). Read through the lens of crypto-assets, these map cleanly onto three primitives: \emph{transactions} (signed state transitions), \emph{native/non-native tokens} (value as protocol state), and the \emph{blockchain} (the shared, append-only history that orders and authenticates those transitions). In the following, we walk through this specification, in turn showing both why it works and where permissionless, pseudonymous governance deviates from the ISO assumptions.

\subsubsection{Transactions as ``records''}

In blockchain-native terms: transactions are the records (signed, ordered state transitions, later anchored by block inclusion); coins/tokens are the information assets (value represented by UTXOs or account balances); and the blockchain is the aggregation (a replicated, append-only ledger that organises those records). Read this way, a confirmed transaction has the core properties of an ``authoritative record'': authenticity (signature/key control), reliability (consensus-confirmed ordering), integrity (tamper-evidence via hashing/chain reorg cost), and usability (public verifiability and traceability) \citep[see][]{Lemieux2016Jul,Wang2021Jul}.

The mapping is not perfect. ISO 15489 implicitly assumes gatekeeping: authorized staff create records, and identities are administratively verified. Depending on the consensus mechanism, a blockchain may fit into this definition (e.g. proof-of-authority and (partially) proof-of-stake) or deviate from it (e.g. proof-of-work). Those permissionless chains depart from the definition as anyone capable of producing a valid signature can write to the ledger, and ``identity'' (unless linked off-chain) is usually just key ownership-pseudonymous \citep{Liang2024Jul}. The mismatch is therefore about governance and attribution, not about transactions failing to function as records.

\subsubsection{Crypto-assets as ``information asset''}

ISO 15489 frames records as both evidence and information assets. Crypto-assets make the ``information asset'' part unusually literal: a coin/token is value instantiated as data, a protocol-defined state entry (UTXO, account balance, token ID) that is created and updated only via valid, consensus-accepted transactions \citep{Xu2023Jul}. A physical substrate is neither present nor required; the value of an asset is grounded in the authoritative history of issuance and transfers within the ledger.

What is distinctive is self-reference. For most information assets, the record points to something else (a licence to use software, a deed to property). For crypto-assets, the ownership record \emph{is} the asset \citep{Wyczik2025May}: control is key control, and transfer is an on-chain state transition.\footnote{Recall the mantra, ``your keys, your coins; not your keys, not your coins'' \citep[cited in][]{levitin2022not}, by Andreas Antonopoulos.} This reflexivity matters for disposition: ``destroying'' an asset usually means making its state unspendable (e.g., burn or key loss) while the underlying record persists.

\subsubsection{Blockchains as ``aggregations''}

ISO 15489 uses \emph{aggregation} for structured set of records. In crypto-assets terms, transactions aggregate into blocks, making a blockchain a structured sets of records: a shared, append-only history defining a system's current state by providing a univocal ordering of transactions. A blockchain's primary purpose is preventing double-spending by enforcing one unambiguous and verifiable transaction history \citep{nakamoto2008btc}.
Unlike enterprise records systems, permissionless ledgers have no designated custodians \citep{Barbereau2023Jul}. Every full node stores (and can independently validate) the aggregation, while inclusion and conflict resolution are handled by protocol rules and consensus rather than administrative control. As a result, lifecycle transitions emerge from network behaviour, not from formal record-keeping decisions.\\

%In sum, crypto mostly repackages familiar information-management roles in a new stack: transactions are records, tokens are information assets, and the chain is the aggregation. The novelty is the enforcement layer (signatures, distributed consensus, and append-only replication) and not the underlying conceptual mapping.

\subsection{Conceptualisation: seven stage document lifecycle for blockchain data}
The specification demonstrates that blockchains can be analysed as records management systems. Next, we conceptualise the seven-stage records lifecycle framework for blockchain data. We consider that blockchain data exhibits characteristics that reshape how each lifecycle stage operates. The following subsections examine each stage sequentially, analysing how protocol-determined mechanisms, append-only architectures, and the on-chain/off-chain boundary strain traditional records management assumptions whilst preserving the fundamental lifecycle structure.

\subsubsection{Creation}
\label{subsubsec:creation}
Traditional document creation assumes an authorised ``creator'' \citep{iso15489} who produces records with accurate and complete metadata for future retrieval. In blockchain systems, creation occurs when a transaction is broadcast to the memory pool and validated through protocol-determined mechanisms. The creator's authority derives from cryptographic proof and protocol compliance rather than institutional delegation.

This distinction matters for governance. Traditional records management employs \emph{administered} governance: policies define who may create records and under what circumstances. Decentralised systems use \emph{embedded} governance (see also, the ``rule of code'' in \citet{DeFilippi2018}): the consensus mechanism validates transactions when protocol requirements enforced by nodes are satisfied.

\subsubsection{Review \& Approval}
\label{subsubsec:review}
Traditional review and approval involves designated reviewers (e.g., notaries, managers, legal teams) who ensure accuracy, legal validity, and compliance \citep{iso15489}. In blockchain systems, these functions are performed distributively through consensus mechanisms \citep{Xu2023Jul}. Review occurs without designated reviewers; validation is algorithmic and collective. Transactions are approved when included in a block that nodes accept as valid.

For permissioned blockchains, traditional review processes may coexist with protocol validation (e.g. Ripple, USDT, Hyperledger). These represent hybrid governance: institutional review followed by protocol validation, where the on-chain/off-chain boundary separates two distinct approval regimes.

\subsubsection{Publication}
\label{subsubsec:publication}
Publication in traditional records management makes approved documents available through appropriate channels \citep{iso15489}. In blockchain systems, publication occurs through peer-to-peer broadcast and block confirmation. A transaction enters publication when a validator picks it from the memory pool and includes it in a valid block.
The transition from review to publication is protocol-determined. While traditional systems allow discretion about timing and channels, blockchain publication is automatic once block validation succeeds.

However, what users experience as crypto-asset publication may occur partially off-chain, for instance when funds are deposited on exchanges, wrapped in smart contracts, or moved to Layer 2 solutions like the Lightning Network. The latter only publishes channel opening and closing states on-chain, whereas intermediate transactions remain in secondary systems with different publication characteristics.

\subsubsection{Active Use}
\label{subsubsec:use}
Active use in traditional records management involves retrieval, viewing, and collaboration, supported by check-out procedures, usage tracking, and access controls \citep{iso15489}.
In blockchain systems, transactions are in active use when they represent spendable value. A Bitcoin UTXO is in active use by being unspent and accessible to the key holder (see Appendix \ref{app:btc}). Transactional atomicity (a transaction either executes completely or not at all) is central: in crypto-asset systems, any state change within a transaction is atomic.

Privacy-enhancing technologies most significantly affect this stage. Mixers, zero-knowledge proofs, and privacy-focused protocols deliberately obscure relationships between transactions \citep{barbereau2023beyond}. From a records management perspective, these technologies interfere with usability (the ISO 15489 requirement that records be connected to the transactions that produced them) by severing, obfuscating, or shielding links between inputs and outputs. This observation is reflected in forensic analysis of, for example, Monero \citep{Koerhuis2020Jun}.

Centralised intermediaries and Layer 2 solutions also interfere with usability. When users hold crypto-assets on an exchange or within a Lightning Network channel, blockchain analysis can trace funds to deposit addresses or channel openings, but cannot observe how users employ their holdings within these systems. This is a recognised challenge for investigators and prosecutors \citep{vanRoomen2025Nov}.

\subsubsection{Revision}
\label{subsubsec:revision}
Immutability, one of blockchain's most cited properties, creates an apparent problem for revision. If confirmed transaction records cannot be altered, how can revision occur? Setting aside edge cases of \emph{record revision} such as successful 51\% attacks reorganising the ledger (e.g. FIRO in January 2021 \citep{firo2021}), or hard forks overriding consensus (e.g. Ethereum's DAO recovery \citep{dupont2017experiments}), the closest equivalent to revision is \emph{state revision}. Individual transaction records are immutable, but the \emph{state} they collectively represent (which keys control what value) changes continuously through new transactions. When coins are spent, new transactions create new UTXOs, effectively revising ownership records whilst preserving complete prior history.

This pattern (append-only revision through supersession rather than modification) is not unique to blockchains. Traditional records management recognises superseding records that replace prior versions whilst maintaining version history~\citep{Kastenhofer2015Sep}. What distinguishes blockchain revision is its \emph{mandatory comprehensiveness}: every revision automatically preserves complete history, whereas traditional systems may optionally delete superseded versions.\footnote{In several blockchains, pruned nodes store only current state rather than full transaction history. However, this is a node-level optimisation that does not affect the network's retention of complete history, as full nodes maintain the entire ledger. Moreover, pruned nodes have operational trade-offs such as reduced ability to verify historical transactions independently or run Layer 2 nodes.}

This invites comparison with Rai stones, large stone discs used as currency by Yap islanders in Micronesia, where ownership is tracked through oral history rather than physical transfer~\citep{Fitzpatrick2020Jan}. Like Rai stones, blockchains maintain a public ledger of ownership changes; unlike Rai stones, blockchain ledgers are cryptographically secured and globally distributed. The principle that authoritative revision of ownership matters more than physical possession is ancient, finding novel implementation in blockchain systems.

The implications for illicit activity analysis are significant. In traditional records management, revision may obscure prior states if version history is not maintained. In blockchain systems, on-chain revision is inherently auditable: every ownership change is permanently recorded. This auditability applies only to on-chain state, but its efficacy explains why most blockchain analysis focuses on follow-the-money approaches \citep{Koerhuis2020Jun,DUDANI2023301576,Frowis2020Jun}.

\subsubsection{Retention}
\label{subsubsec:retention}
Traditional records management treats retention as a governed stage: retention schedules specify how long different record types must be maintained before disposition. Responsibility is assigned, retention periods defined, and costs budgeted \citep{iso15489}.
In blockchain systems, retention is the default state. All confirmed transactions are retained indefinitely by every full node. This inversion from selective to comprehensive retention has several consequences.

First, traditional concerns about inadequate retention (records destroyed before their retention period expires) are replaced by concerns about excessive retention (records that cannot be disposed of even when no longer required). Second, responsibility for retention is distributed rather than assigned; no single entity bears custodial responsibility. Third, retention costs are borne collectively through storage requirements for full nodes, creating incentive structures that may eventually affect retention comprehensiveness.

Comprehensive retention has implications for legal and regulatory frameworks premised on data destruction. For instance, the European General Data Protection Regulation's right to be forgotten, assuming an address or transaction is linkable to a natural person's identity, conflicts with blockchain immutability \citep[see][]{Tatar2020Sep}. Lifecycle analysis helps clarify but does not resolve this tension.

Off-chain data does not share this comprehensive retention. Layer 2 transactions, exchange records, and NFT metadata stored on IPFS are subject to different retention mechanisms and may be lost if not actively maintained. The on-chain/off-chain boundary thus creates asymmetric retention guarantees within what users perceive as a single system \citep{Kaisto2024Sep}.

\subsubsection{Disposition}
\label{subsubsec:disposition}
Record persistence through all forms of disposition distinguishes blockchain systems fundamentally from traditional records management, straining the lifecycle analogy most at ``disposition''. ISO 15489 defines disposition as retiring records through ``archival or secure destruction'' \citep{iso15489}. Blockchain transaction records are never destroyed; they persist indefinitely across distributed nodes. 

In a blockchain system, the disposition lifecycle stage applies differently to the assets and their documentation: it affects the information asset (i.e. the crypto-asset unit) without affecting the record (i.e. the transaction). Disposition of crypto-asset units can take four distinct forms:
\begin{itemize}
    \item \textbf{Functional inaccessibility:} When private keys are accidentally lost or deleted/destroyed, corresponding assets become permanently inaccessible. Transaction records persist, but the information assets they document are effectively disposed of, removed from circulation without authoritative retirement. A substantial amount of Bitcoin is permanently inaccessible due to key loss, wallet errors, and forgotten holdings \citep{Khatib2025Mar}. This resembles accidental destruction of physical records rather than deliberate disposition.
    \item \textbf{Intentional burning:} Some blockchain systems include mechanisms for deliberately destroying assets. Sending tokens to verified burn addresses (e.g., 0x000... 000) or executing smart contract burn functions constitutes intentional disposition. Unlike traditional secure destruction, burned assets leave permanent records of their destruction: \emph{auditable destruction}.
    \item \textbf{Protocol-level obsolescence:} Protocol upgrades or chain migrations may render prior records obsolete. When Tether migrated from Omni to Ethereum, transactions on the original chain became historical artefacts rather than active records \citep[see][]{Lyons2023Mar}. This disposition is neither accidental nor individually intentional but emerges from collective decisions about protocol evolution.
\end{itemize}

\subsection{Application}
The three case studies illustrate how these principles manifest for Bitcoin, fungible tokens (Tether), and non-fungible tokens respectively. Each demonstrates the seven lifecycle stages whilst highlighting the characteristics discussed here. Note, as with the example of notable implementations of privacy-enhancing technologies, there can be overlaps between lifecycle stages. 

\subsubsection{Bitcoin}
\label{app:btc}

In a novel example, let us now follow 1 bitcoin throughout the same seven-stage records lifecycle framework (based on the original whitepaper of \citet{nakamoto2008btc} and \textit{The Book of Satoshi} \citep{satoshibook2014}:
\begin{enumerate}
    \item \textbf{Creation}: An asset (UTXO) on the Bitcoin blockchain is created either through a coinbase transaction that generates a newly mined UTXO or the use of an existing UTXO as an input to a user transaction that reallocates UTXOs on the output side.  In both cases, the transaction record includes metadata (block height, timestamp, miner fees).
    \item \textbf{Review \& Approval}: The Bitcoin network nodes validate the block by verifying the proof-of-work, checking that the transaction data follows protocol rules, and ensuring no double-spending has occurred. Consensus is reached when the majority of nodes accept the block as valid.
    \item \textbf{Publication}: The validated block is broadcast across the peer-to-peer network and appended to the blockchain. The transaction record becomes publicly accessible through block explorers, and the bitcoin is now spendable.
    \item \textbf{Active Use}: The published UTXO may serve as a store of value, a proof of assets, or collateral that can be verified by any network participant in real time.
    \item \textbf{Revision}: Whilst UTXOs cannot be altered due to blockchain immutability, the ownership record may change with a change of private key control off-chain. When spent, a new UTXO is created, effectively revising the record of who controls the bitcoin whilst preserving the complete transaction history.
    \item \textbf{Retention}: The transaction records are permanently maintained across thousands of distributed nodes worldwide. The blockchain protocol ensures indefinite retention, with each full node storing the complete history to maintain network integrity and enable verification of all future transactions.
    \item \textbf{Disposition}: UTXOs cease to exist when they are spent. Their history, however, remains permanently archived in the blockchain. UTXOs may also be disposed of by destroying the associated private key(s).  Some deliberately send bitcoin to provably unspendable addresses, creating a permanent but unusable record, or burning the coins into fees that will be earned by the miners of the next block \citep[see][]{Khatib2025Mar}.
\end{enumerate}

\subsubsection{Ethereum Virtual Machine-based}
\label{app:ethereum}
The seven-stage document lifecycle framework can be applied in the context of EVM-based blockchains for fungible and non-fungible tokens. 

\paragraph{Fungible token.}
\label{app:usdt}
The following considers 1 USDT, a fungible token following the ERC-20 standard on an EVM blockchain, from the seven-stage lifecycle framework:

\begin{enumerate}
    \item \textbf{Creation}: A user deposits \$1 USD into Tether Limited's bank account. Tether's system creates a minting transaction record with metadata (user ID, deposit amount, timestamp, bank reference) and issues a smart contract instruction to mint 1 USDT token on the Ethereum blockchain.
    \item \textbf{Review \& Approval}: Tether's compliance team verifies the bank deposit, confirms KYC/AML requirements are met, checks that reserves are sufficient to back the new token, and ensures the minting request complies with regulatory standards before authorizing the smart contract execution.
    \item \textbf{Publication}: The approved minting transaction is executed on the Ethereum blockchain, creating 1 USDT in the user's wallet address. The transaction is recorded on-chain with a unique transaction hash and becomes publicly viewable on blockchain explorers, whilst Tether updates its treasury dashboard showing total supply.
    \item \textbf{Active Use}: The user transfers the USDT to a crypto-asset exchange to trade, then to a DeFi protocol for lending, and later to another individual for payment. Each transfer is recorded on the blockchain with complete transparency, and the token's authenticity can be verified in real time through the smart contract.
    \item \textbf{Revision}: The USDT ERC-20 contract\footnote{https://etherscan.io/token/0xdac17f958d2ee523a2206206994597c13d831ec7} was deployed in November 2017 by Bitfinex/Tether. Whilst the USDT contract has remained unchanged since launch, its source code contains a function which can be used if the contract is ever upgraded to a new version. If used the function would act as a proxy forwarding smart contract calls to a new implementation.\footnote{That is a \texttt{deprecate()} function that would redirect contract calls to an \texttt{upgradedAddress} if called} The migration of USDT from Omni to Ethereum mentioned in Section \ref{subsubsec:disposition} was not a smart contract upgrade, but a cross-chain migration where the token was disposed of on Omni and created \textit{ex novo} on Ethereum.
    \item \textbf{Retention}: Transaction records remain permanently stored on the blockchain across distributed nodes. Tether maintains parallel financial records of fiat reserves, user deposits, and minting/burning events for the legally required period (typically seven years) to satisfy regulatory audits and compliance requirements.
    \item \textbf{Disposition}: When a user redeems 1 USDT for \$1 USD, Tether executes a 'burn' transaction that permanently destroys the token through the smart contract. The burning event is recorded on-chain as the final transaction, the fiat is returned to the user's bank account, and whilst the complete transaction history remains in the blockchain archive, the specific USDT token ceases to exist and is removed from circulating supply.
\end{enumerate}

\paragraph{Non-fungible token.}
\label{app:nft}

The following considers a (fictional) non-fungible token, following the ERC-721 standard on an EVM blockchain, from the seven-stage lifecycle framework:

\begin{enumerate}
    \item \textbf{Creation}: An artist creates an artwork and uploads it to IPFS (InterPlanetary File System). They then mint an NFT using an ERC-721 smart contract on an EVM blockchain, embedding metadata (title, description, creator wallet address, IPFS hash, timestamp, royalty percentage) that links to the artwork and establishes provenance.
    \item \textbf{Review \& Approval}: An NFT marketplace reviews the submission for compliance with platform policies, verifies the creator's identity through wallet signatures, checks that the artwork does not infringe copyright or contain prohibited content, and ensures the smart contract metadata meets technical standards before listing approval.
    \item \textbf{Publication}: The approved NFT is listed on the marketplace with its metadata publicly visible. The minting transaction is recorded on the Ethereum blockchain, the token receives a unique identifier (token ID), and the artwork becomes discoverable through the platform's interface and blockchain explorers.
    \item \textbf{Active Use}: A collector purchases the NFT, displaying it in their digital gallery and virtual reality exhibition. They later loan it to a metaverse museum for a temporary showcase. Each transfer and interaction is recorded on-chain, with the NFT's ownership and exhibition history verifiable in real time through the blockchain.
    \item \textbf{Revision}: The artist discovers an error in the artwork's description metadata. Whilst the on-chain token data is immutable, they update the off-chain metadata file on IPFS, creating a new version with corrected information. The NFT's token URI now points to the updated metadata, whilst the blockchain maintains a complete record of the original minting transaction.
    \item \textbf{Retention}: The NFT's ownership record remains permanently stored on the Ethereum blockchain across thousands of nodes. The associated artwork file persists on IPFS through pinning services, and marketplaces maintain historical sales data, exhibition records, and price history for the legally required period to support provenance verification and tax compliance.
    \item \textbf{Disposition}: The collector decides to `burn' the NFT (perhaps as part of a promotional event or artistic statement \citep{Calvo2024Mar}). They send the token to a verified burn address (0x000...000), permanently removing it from circulation. The burning transaction is recorded on-chain as the final event, the token becomes non-transferable, and whilst the complete ownership history and metadata remain archived in the blockchain, the NFT itself is effectively destroyed and can never be traded again.
\end{enumerate}

\subsection{Note on privacy-enhancing technologies}
\label{subsubsec:privacycoins}
The functional complexity of privacy-enhancing technologies in cryptocurrency protocols derives in part from their capacity to operate across multiple lifecycle stages rather than at a single point. This warrants additional explanation. We do so on the basis of so called `privacy coins' (Monero, Zcash, Litecoin, Firo, Dash, and Bitcoin Cash) mapped against our seven-stage framework.

Monero, which is UTXO-based, employs three mechanisms engaging separate lifecycle stages. Ring Confidential Transactions (RingCT) blind transaction amounts via Pedersen commitments \citep{Noether2015} at Creation (stage 1), such that no unblinded record is ever produced. Stealth addresses operate at Publication (stage 3), generating one-time addresses per transaction at broadcast to ensure recipient unlinkability. Ring signatures engage Retention (stage 6) in a structurally novel manner: settled transaction outputs are retained as active decoys for future transactions, extending their functional utility indefinitely beyond their own settlement.
More recently, Monero is planning to upgrade to Full Chain Membership Proofs++ (FCMP++) \citep[see][]{parker2026}. In FCMP++, ring signatures will be replaced with a zero-knowledge membership proof that proves "I know the secret key for some output in the entire UTXO set"\footnote{The anonymity set grows from 16 explicitly selected outputs to a pool of millions of outputs.}, without naming any decoys. FCMP++ membership proofs still engage at stage 6. Retention, but in a transformed manner: all unspent outputs are implicitly enrolled as anonymity-set members at proof construction, such that no explicit decoy sampling occurs.

Zcash, originally a fork of Bitcoin and thus also UTXO-based, employs zk-SNARKs to constitute the complete transaction record \citep{Akcora2022Jan} at Creation (stage 1), such that no prior unblinded record exists. This architecture collapses Review \& Approval (stage 2) into Creation (stage 1), as mathematical proof substitutes for human compliance verification. These properties apply exclusively to shielded transactions, as transparent Zcash transactions carry no privacy properties.

Litecoin's MimbleWimble Extension Blocks (MWEB) engage three lifecycle stages \citep{Silveira2024Aug,montalto2026}. Pedersen commitments blind transaction amounts at Creation (stage 1). A cut-through mechanism prunes intermediate transaction outputs by collapsing transaction chains into their net result, mapping onto Revision (stage 5) as a retrospective transformation of existing records while preserving verifiable net state. The pruned intermediate records are subsequently permanently retired, engaging Disposition (stage 7).
%MWEB is the only protocol discussed with a meaningful Revision stage engagement, owing to the retrospective nature of cut-through operating on already-settled records.

Firo's Lelantus Spark protocol \citep{Jivanyan2023Jul} engages three lifecycle stages. At Creation (stage 1), a new anonymous record is minted from a cryptographic commitment. Spark one-time addresses and diversified recipient keys provide publication-layer unlinkability at Publication (stage 3), analogous to Monero's stealth addresses. Firo's burn-and-redeem mechanism mandates Disposition (stage 7) as a prerequisite for the creation of any private record: the originating transparent coin record must be cryptographically retired before the private record can be issued.

Dash employs PrivateSend, a CoinJoin-based coordination mechanism that produces only an aggregated transaction record \citep{Deuber2021Oct} at 1. Creation, with individual transaction intents never recorded on-chain. Unlike cryptographic approaches, PrivateSend reintroduces a meaningful Review \& Approval stage (2) through the multi-party coordination required before broadcast. Privacy obfuscation exists only at Publication (stage {3}), making it coordinative rather than cryptographic, and opt-in only.

Bitcoin Cash employs CashFusion, a more sophisticated CoinJoin variant that improves resistance to linkability analysis through a more complex combinatorial mixing algorithm, but shares the same fundamental lifecycle logic as Dash's PrivateSend across Creation (stage 1), Review \& Approval (stage 2), and Publication (stage 3).

Across all protocols examined, no privacy-enhancing technology meaningfully engages the Active Use stage (4). This represents a structural gap between the records lifecycle framework and blockchain transaction models, as post-settlement query privacy (for instance, data leakage through light client protocols) constitutes a separate class of privacy concern unaddressed by any of the protocols discussed.
\section{Discussion}
\label{sec:discussion}

%% TOM: I drafted a discussion section based on the expected findings and our discussion of yesterday. Feel free to go thourhg and edit if time allows :)

We have specified blockchain from the perspective of ISO 15489 and, thereupon, have conceptualised and applied the seve-stage document lifecycle framework. Blockchain's distinctive features do not invalidate the lifecycle framework but require recognising that stages represent \emph{states of data} rather than \emph{administrative decisions about data}. Creation is protocol-determined rather than institutionally authorised. Review and approval occur through collective validation rather than designated reviewers. Publication is automatic upon consensus rather than discretionary. Active use is atomic and defined by spendability rather than administrative classification. Revision operates through append-only supersession rather than modification. Retention is comprehensive by default rather than governed by schedules. Disposition affects assets without destroying records.
The on-chain/off-chain boundary creates consistent complications across all stages of the framework: centralised intermediaries, Layer 2 solutions, and off-chain metadata may follow different lifecycle patterns than on-chain data. Privacy-enhancing technologies similarly complicate lifecycle visibility, particularly during active use.
Re-framing crypto-assets as records has both theoretical and practical implications.

\subsection{Contributions to theory}

First, the conceptualised seven-stage lifecycle framework for blockchain data challenges implicit assumptions of clean, discrete stages. Much of the existing literature on crypto-assets and blockchain treats stages such as creation, publication, use, and disposition as analytically separable. However, in reality, these stages directly overlap and are linked into one another. For example, using crypto-assets as collateral in the Active Use stage may imply changeing key control temporarily (stage 5. Revision). The lifecycle framing thus foregrounds the need for theories that accommodate entanglement, feedback loops, and non-linear progression, rather than assuming orderly transitions between stages.

Second, the framework for blockchain data problematizes normative assumptions about crypto-asset use over time. Prevailing research often evaluates crypto-assets based on current or intended uses (be that using heuristic methods, or studies of specific smart contract functionalities), implicitly assuming stability in how these assets function as technology. The lifecycle lens highlights that data artifacts deemed legitimate or socially beneficial at one point may later be repurposed in harmful, illicit ways -- or vice versa (as also discussed in \citet{Barbereau2023Jul,Boss2025Jul}). This raises theoretical questions about responsibility, temporality, and moral accountability: who bears responsibility for downstream misuse, and at what stage? What constitutes misuse? By emphasizing longitudinal trajectories rather than static snapshots, the lifecycle framing encourages theorising around shifting value, risk, power, and legitimacy across time. 

Third, the conceptualization provides a meta-structure for organizing and integrating fragmented research streams. Crypto-asset scholarship is often siloed around specific implications -- technical efficiency, market behaviour, regulation, or social impact. By situating these contributions within a shared lifecycle framework, researchers can more clearly articulate where their work intervenes, how insights from one stage condition outcomes in another, and where theoretical blind spots remain. In this sense, the lifecycle perspective functions as a unifying theoretical scaffold rather than a competing explanatory theory.

\subsection{Implications for practice}

Practical implications for authorities investigating, regulating, or supervising crypto-asset-related activity are twofold. First, treating crypto-assets as lifecycle-bound data enables practitioners to decompose complex problems into stage-specific challenges. Rather than addressing failures only at the point of use -- by following the money -- law enforcement agencies and others may trace abuse or misuse back to earlier lifecycle stages in the framework, be it at the stage of creation or disposition. For example, considering seized crypto-assets that are to be liquidated by the competent authority, the problem of future use of (formerly) illicit assets arises. The German Federal State of North-Rhine Westphalia auctioned off seized Bitcoin in 2021 with some auctions closing well above market prices \citep{zeit_bitcoin2021}. While a bidder’s willingness to pay may have many different motives, at least one successful bidder was found to commingle the auctioned funds with proceeds from a sanctioned entity -- resulting in the suspicion of using the funds, seemingly cleaned through the auction, to whitewash other proceeds. Understanding crypto-assets in their respective stage in the lifecycle would have foregrounded future-oriented responsibility in this case. Practitioners are prompted to consider not only whether a crypto-asset is used appropriately today, but how it might be misused tomorrow and what safeguards exist to mitigate such risks. This could include anticipating secondary markets (as part of stage 4. Use), unintended data persistence (stage 6. Retention), and the absence of clear disposal, decommissioning or whitelisting mechanisms (stage 7. Disposition).

The framework highlights gaps in existing operational and regulatory practices, particularly around end-of-life considerations. While significant attention is devoted to creation and use, comparatively little guidance exists on how crypto-asset-related data should be sunset, archived, or rendered inert when it no longer serves its intended purpose. Though reference cases exist (see for example in Hesse, Germany \citep{delCastillo2024Mar}), recognizing disposal as a legitimate lifecycle stage draws attention to unresolved practical questions around data permanence, legal liability, and environmental and social externalities.

Second, the lifecycle approach may facilitate tracing in investigative work related to crypto-assets. The lifecycle approach shifts the focus of the investigators from “follow the money” to “change of ownership” -- who can be determined to have been the owner of a UTXO at the time of creation and at the time of disposition or destruction. Thus, a chain of transactions should be seen as a sequence of lifecycles with potentially different owners. This ensures consistency of ownership attribution along a chain of transactions, increasing the evidential value of blockchain investigations and providing a conceptual framework for argumentation schemes \citep[see][]{fintech3020014}.

A third practical implication concerns data protection compliance. The tension between blockchain immutability and the GDPR right to erasure is well-documented \citep{Tatar2020Sep}, but the lifecycle lens reframes it productively: rather than treating immutability as a binary obstacle, practitioners can intervene at earlier stages to limit what reaches the chain in the first place. At stages 1. Creation and 2. Review \& Approval, design choices -- storing only cryptographic commitments on-chain, keeping personal data off-chain -- can structurally pre-empt erasure conflicts before they arise. At stage 6. Retention, the on-chain/off-chain asymmetry noted in Section~\ref{subsubsec:retention} matters practically: deletion of off-chain identifiers and exchange records can meaningfully degrade re-identification risk even where the underlying transaction record persists.
Lifecycle-aware compliance should therefore focus regulatory intervention on the stages where personal data first enters -- or can be prevented from entering -- the chain.

\subsection{Limitations}
% TOM TO EDIT BASED ON UPDATED METHOD

This work is conceptual in nature and therefore subject to several limitations \citep{mora2008conceptual,Jaakkola2020conceptual}. First, the lifecycle framework is not empirically validated within this article. While it is grounded in existing literature and illustrative examples (Sections \ref{app:btc} and \ref{app:ethereum}), future empirical studies are needed to assess its descriptive accuracy and explanatory power across different crypto-asset systems and contexts. Second, in abstracting crypto-assets as data that move through a lifecycle, the framework necessarily simplifies socio-technical complexity, potentially overlooking context-specific dynamics such as technical nuances in protocols, jurisdictional variation, or emergent privacy methods. Finally, the proposed lifecycle stages are presented as an analytical heuristic rather than as discrete or exhaustive categories; boundaries between stages may be porous or contested in practice. These limitations underscore the need for subsequent empirical and comparative work to refine, extend, and operationalize the framework. Each presents avenues for future research.
\section{Outlook}
\label{sec:outlook}

The seven-stage records lifecycle framework for blockchain data advanced in this work opens several avenues for future research and practice. Empirically, scholars can examine how its stages intersect, overlap, or collapse in different contexts or ledgers, thereby testing the robustness of the framework. Conceptually, further work is needed to theorize responsibility, accountability, and governance across time, particularly as crypto-assets evolve beyond their originally intended purposes. For practice, the framework invites the development of lifecycle-aware tools, standards, and policies that anticipate not only use, but also misuse, adaptation, and end-of-life considerations. More broadly, viewing crypto-assets as data with temporal trajectories may inform the study of other emerging digital artifacts characterized by persistence, recombination, and uncertain futures.

\section*{Funding}
This research did not receive any specific grant from funding agencies in the public, commercial, or not-for-profit sectors.

\section*{Declaration of Conflicting Interests}
The author(s) declared no potential conflicts of interest with respect to the research, authorship, and/or publication of this article.

\newpage
\bibliography{references}

%\appendix
%\input{sections/99_appendix}

\end{document}